\documentstyle[prl,aps, multicol]{revtex}

\newcommand{\be}{\begin{equation}}
\newcommand{\ee}{\end{equation}}
\newcommand{\br}{\begin{eqnarray}}
\newcommand{\er}{\end{eqnarray}}

\newcommand{\bd}{\begin{displaymath}}
\newcommand{\ed}{\end{displaymath}}

\newcommand{\bfig}{\begin{figure}}
\newcommand{\efig}{\end{figure}}

\def\3cdot{\cdot \cdot \cdot}

\def\om0{\omega _0}
\def\Om0{\Omega _0}

\def\text#1{{\rm{#1}}}

\def\->{\rightarrow}
\def\=>{\Rightarrow}
\def\-->{\longrightarrow}
\def\==>{\Longrightarrow}

\def\pr{^\prime}
\def\pr2{^{\prime\prime}}

\def\bfig{\begin{figure}}
\def\efig{\end{figure}}

\begin{document}
\draft
\title{Quantum computation with mesoscopic superposition states}
\author{M. C. de Oliveira\cite{marcos} and W. J. Munro}
\address{Centre for Laser Science, Department of Physics, \\
University of Queensland, QLD 4072, Brisbane, Australia.}
\date{\today}

\maketitle
\begin{abstract}
We present a strategy to engineer a simple cavity-QED two-bit universal quantum
gate using  mesoscopic distinct quantum superposition states. The dissipative
effect on decoherence and amplitude damping of the quantum bits are analyzed and
the critical parameters are presented.
\end{abstract}

\vspace{5mm}


\section{Introduction}
%
Quantum Mechanics is now fundamental to the modern world we live
and interact with, not being just the abstract realm of theoretical physics.
Many new areas of emerging technology depend on the principles contained
within it \cite{dec}. One of the most striking features of 
quantum systems are superposition states. 
They have given rise to a large amount of discussion in the literature
\cite{despagnat} and now play a central role for the
recent developments made in the area of quantum
information. This is due to their possibility to encode
information in a way impossible to be attained by any classical system.
Quantum computation has become a significant subject within quantum information theory,
due to the powerful property of superposition states to execute large
parallel processing. Quantum information research has also
improved significantly the understanding about the quantum systems involved on
the factual realisation of a quantum computer
and has raised many interesting 
problems such as in the encoding of information
\cite{inf}, entanglement of states \cite{entang} and quantum cryptography
\cite{cryp}.

A number of core technologies are currently under investigation for
constructing a quantum computer which is necessary to  fully implement 
quantum algorithms. These
include ion-traps\cite{wineland,wineland2}, cavity QED\cite{sleator,domokos,nogues}, solid 
state\cite{kane,vrijen} and 
liquid state NMR\cite{chuang} to name but a few.
The proposals to engineer a quantum computer or as a first step a single
logic gate in the realm of quantum optics are generally based on discrete
atomic states and cavity field number states of zero and one photons.
A central proposal which has gained much attention in recent years is 
the Cirac and
Zoller \cite{zoller} trapped ions scheme to encode a n-conditional gate.
We also cite the proposals of Sleator and Weinfurter \cite{sleator} 
and of Domokos {\it
et al.}\cite{domokos} based on cavity-QED (quantum electrodynamics) 
technology and dealing with
two-bit universal gates. Experimentally, there are few initiatives for logical
operations in ion traps \cite{wineland2} and in NMR \cite{chuang}, which 
 allow for a scalable implementation.  These proposals require
a technological domain, which to date  has not been attained
\cite{wineland,haroche,kimble}. In cavity-QED technology, for optical frequencies, a
conditional interaction between two-modes,  the idler and pump, 
 have been proposed to encode a phase gate (P-gate) \cite{kimble} due to the  
high non-linearity that can be presented by single atoms. At microwave frequencies, logical 
elements  have been demonstrated experimentally as a means 
of  encoding a quantum memory with a single photon \cite{memdavid}.

In this article we are going to focus on cavity QED and the technology
associated with it. Cavity QED has had a very rich past and has been
instrumental in a huge amount of fundamental quantum and atom optical
research \cite{harochept}. Such a system has been used for  photon number  quantum non demolition measurements
\cite{nogues,brune}, generation of single Fock state \cite{nogues,numb} and generation and
measurement of the time of decoherence of Schr\"odinger cats states
\cite{brune,brune2,raimond}. With such a rich history recent attention in cavity QED has 
been focused on quantum information. With the non demolition measurement of a single photon
number in the cavity \cite{nogues,numb}, the technology became 
available to encode qubits and realise a quantum gate \cite{nogues}. The quantum 
information proposals based on
cavity QED technology \cite{sleator,domokos} makes use of only zero and one
field number states. More recently there is the significant evidence 
of generation of trapped states of
more than one photon \cite{varcoe} which could be used in an encoding
scheme.

With a {\sc CNot} gate based on an encoding scheme using zero and one Fock
states, spontaneous errors have a disastrous effect. Quantum
information is irreversibly lost. It is possible to protect  the system against
such errors. In fact to protect the qubit against general one qubit errors it
is necessary to encode the original state by distributing  its quantum information
over at least five qubits.  Basically the 5-qubit quantum circuit takes the initial state with four extra qubits in the state
$\left|0\right\rangle$ to  an encoded state. This state is then protected versus
 all single qubit errors. Decoding this state and then applying a simple
unitary transformation yields the original state. Implementing a
five qubit error correcting code is quite expensive in terms of quantum
resources. Other encoding schemes may allow simpler error correction circuits.

There is no fundamental reason to restrict oneself to 
physical systems with
two dimensional Hilbert spaces for the encoding. It may be more natural in
some contexts to encode logical states as a superposition over a large
number of basis states. Significant advances can be achieved. For instance
in the protection against errors incoming due to the coupling of the qubit
system to a dissipative environment. Recent work by
Cochrane {\it et. al.}\cite{munro} have proposed how macroscopically
distinct quantum superposition states (Schr\"odinger cat states) may be used as
logical qubit encoding.  
Spontaneous emission causes a bit-flip error in these superposition state qubit
encoding, which is easily
corrected by  a standard 3-qubit error correction circuit (compared 
to five qubits for Fock states).
This is particularly relevant, as the bit-flip error is much easier to fix
than spontaneous emission errors in Fock state systems.
Another good reason for using superposition of coherent states to 
encode qubits, is
that they are naturally generated in any cavity system, while number states of
more than one photon require a large amount of control \cite{varcoe}.

In this paper we propose how even and odd mesoscopic coherent 
superposition of states
can be used to implement and encode a {\sc CNot} quantum gate in a 
realistic superconducting
cavity-QED system, where those states were already generated \cite{harochept,brune2}. 
We define the even cat state as the $0$ qubit and the odd cat state
as the $1$ qubit. This encoding can be represented as
\begin{eqnarray}
\label{a1}
\left| 0\right\rangle_{L} &\equiv&\frac{1}{N_{+}}\left( \left|
\alpha \right\rangle +\left| -\alpha \right\rangle \right) ,\\
\label{a2}
\left|1\right\rangle_{L} &\equiv&\frac{1}{N_{-}}\left( \left| \alpha 
\right\rangle
-\left| -\alpha \right\rangle \right) .
\end{eqnarray}
where $N_{\pm}=\sqrt{2(1\pm e^{-2\left|\alpha\right|^{2}})}$. This normalisation
is important and will be retained throughout the paper.

Given the generation of the two logic qubits how does one implement a
quantum gate in cavity QED. Essentially any two-bit quantum gate is 
universal \cite{sleator,barenco}.
One of these universal quantum gates is the control not gate ({\sc CNot}) 
and consist of a conditional
gate - here if the control bit is 0 the target bit will be 
maintained, but if the control
bit is 1 the target bit will suffer a flip transform to 0. The {\sc CNot} gate can
be engineered by two  Hadamard transforms\cite{aharonov}  plus a phase (P)
transform \cite{munro,aharonov}. The Hadamard transform is a 
single qubit operation that leads to a rotation in the state  while the 
P-transform is a conditional two-bits transform necessary to identify
the state of the control bit. The question posed here is how to identify these
Hadamard and P transforms in a realizable physical cavity QED system when
the encoding for the qubits is in terms of odd and even cat states.

To begin this paper we show how the apparatus similar to the one used 
to generate Schr\"odinger cat field states \cite{brune,raimond,davidovich} can be generalised to 
perform a {\sc CNot} gate conditional transform involving two levels of a
Rydberg atom and the field
mesoscopic superposition state. Here  the two levels of a Rydberg 
atom are considered to encode the controlled (or target) bit and the field cat state will be the
control bit. Since the generation of Schr\"odinger even and odd cat field states in
cavity QED experiments  is dependent
of a conditional measurement \cite{brune2,raimond}, giving a random outcome, we propose
in Section III a strategy based on resonant atomic feedback \cite{milburn}
which allow us to definitely prepare the state of the control bit. The essence of 
this proposal involves using a feedback scheme based on the injection of appropriately prepared atoms.
Basically the state of the cavity is monitored indirectly via the detection
of atoms that have interacted dispersively with it. If the cavity field state is 
not  in the required state, a photon is injected into the cavity. Finally 
in the last section of this paper we present a reasonable detailed discussion of
dissipation and their effect on the {\sc CNot} gate. We explicitly discuss the
advantages of encoding with superposition states over zero one photon number
states used in previous proposals \cite{sleator,domokos}. Attention is
focused on the decoherence phenomenon, as this is one of the main difficulties
for quantum computation.

%
\section{Superposition State Encoding}
%
In the last few years a great amount of experimental progress 
in  cavity QED has enabled 
work at the level of single atoms and single photons, where only two
electronic energy states of Rydberg atoms participate in the exchange of a
photon with the cavity \cite{harochept}. This has enabled cavity QED technology to be responsible for
a large number of interesting experiments showing, the generation of mesoscopic
coherent superposition field states, called Schr\"odinger cat states
\cite{brune}, the decoherence phenomenon \cite{brune2} and
non-local entanglement of quantum systems \cite{memdavid}. These systems have
gained much 
attention due to the quantum non demolition (QND) property of measurement on the
field photon number by atomic interferometry \cite{brune}.

Our experimental proposal is based on the cavity-QED
scheme\cite{haroche,brune,raimond,davidovich} to 
 generate the field superposition states and is depicted schematically in Fig.(1). It consists
of a Rydberg atom beam crossing three cavities, R\( _{1}^{\phi} \),
 C and R\( _{2}^{\theta} \).  Here R\( _{1}^{\phi} \) and R\( _{2}^{\theta} \) 
 are Ramsey zones and C is a superconducting
Fabry-Perot cavity of high quality factor\cite{paulo}. To achieve our desired encoding the atoms are
initially prepared at B in circular states of quantum principal number of the order
of 50. Such atoms are well  suited for this scheme since their lifetime is
over \( 3\times 10^{-2} \)s \cite{haroche,brune,davidovich}.

 The R\( _{1}^{\phi} \) and R\( _{2}^{\theta} \) cavities, where classical fields
resonant with an atomic \( \left| g\right\rangle  \) \( \rightarrow  \) \( \left|
e\right\rangle  \) transition (51.099 GHz)\cite{numb} are injected
during the time of interaction with the atoms, constitute the usual setup for
Ramsey interferometry \cite{haroche}. There, for a selected atomic velocity, the
state of the atom will suffer a rotation in the
vector space spanned by \( \left\{ \left| e\right\rangle ,\left| g\right\rangle \right\}  \).

The experiment is started when one selects the initial state of an atom prepared
in the \( \left| g\right\rangle  \) or \( \left| e\right\rangle  \) by the laser field L. This atom has a
resonant interaction with the field in R\( _{1}^{\phi} \) given
by\cite{haroche,carolina}
\begin{equation}
\label{a12}
H_{I}=\hbar \Omega \left( a_{r}\sigma ^{+}+a_{r}^{\dagger }\sigma ^{-}\right) 
\end{equation}
where
\( \sigma ^{+}\equiv \left| e\right\rangle \left\langle g\right|  \) and 
\( \sigma ^{-}\equiv \left| g\right\rangle \left\langle e\right|  \)
are the atomic pseudo-spin Pauli operators, \( a_{r}^{\dagger } \) (\( 
a_{r} \))  are the creation (annihilation) operator
for the mode of the field in R\( _{1}^{\phi} \) and \( \Omega \) is the one
photon Rabi frequency. With a proper choice of the field phase \(\phi\) in R\( _{1}^{\phi} \)
the atomic states \( \left| g\right\rangle  \) and \( \left|
e\right\rangle  \) are rotated to

\begin{eqnarray}
\label{a22}
\left| g \right\rangle &\rightarrow& \frac{1}{\sqrt{2}}\left( \left|
g\right\rangle +e^{-i\phi}\left| e\right\rangle \right)
\\
\left| e \right\rangle &\rightarrow& \frac{1}{\sqrt{2}}\left( \left|
e\right\rangle -e^{i\phi}\left| g\right\rangle \right)
\end{eqnarray}

The cavity C is tuned near the resonance of the transitions between the atomic
states \( \left| e\right\rangle  \) and \( \left| i\right\rangle  \), a reference
state corresponding to the higher level from \( \left| e\right\rangle  \).
The frequency of the transition \( \left| e\right\rangle  \) \( \rightarrow  \)\( \left| i\right\rangle  \)
is 48.18  GHz \cite{brune} and is distinct of any transition involving the level \( \left| g\right\rangle  \).
The mode geometry inside the cavity is  configured in such a way that the intensity of the
field rises and decreases smoothly through  with the atomic trajectory inside C. For
sufficiently slow atoms and for sufficiently large cavity mode detuning
from the \( \left| e\right\rangle  \) \( \rightarrow  \)\( \left| i\right\rangle  \)
frequency transition, the atom-field evolution is adiabatic and no photonic
absorption or emission occurs \cite{haroche}. On the other hand, dispersive
effects emerge
- an atom in the state \( \left| e\right\rangle  \) crossing C induces a phase
shift in the cavity field which can be adjusted by a proper selection of the atomic velocity (\( \sim 100 \)
m/s) \cite{brune}.  For a \( \pi  \) phase shift the coherent field 
\( \left| \alpha \right\rangle  \) in C transforms to \( \left| -\alpha \right\rangle  \).
On the other hand, the phase shift caused by an atom in the \( \left| g\right\rangle  \)
state is null. The atom field interaction can be  written effectively as
\cite{carolina2}
\begin{equation}
\label{a3}
H_{off}=\hbar \Omega_{2}  a^{\dagger}a\;\sigma^+\sigma^- 
\end{equation}
where $\Omega_{2}$ is the effective Rabi frequency for the interaction of the atom with
the field and  $a^{\dagger}$ ($a$) is the creation (annihilation) operator for the field in C.
After the atomic interaction with the field in C, the atom crosses the second Ramsey
zone $R^{\theta}_{2}$ which introduces a new rotation in the atomic vector
space, analogously to Eq. (4) and (5), but for the phase $\theta$. The atomic state is detected in 
D  by an ionization zone detector, instantaneously giving the atomic state
and the field state in C.  This is due to the entanglement of their 
states. The important point  we emphasise here is that
 the resonant interaction of the Ramsey zones can be used as Hadamard
transform since they induce rotations in the vector space of the target bit (atomic state)
and the off-resonant interaction between atom and field in C can be 
 used for the P-transform\cite{munro}.

We begin the description of the implementation of the {\sc CNot} gate 
by specifying that the coherent field state will be responsible for the encoding of the control bit 
and the atomic states \( \left| g\right\rangle  \)
and \( \left| e\right\rangle  \)  will be the target bits \( \left| 0\right\rangle_{T}  \)
and \( \left|1\right\rangle_{T}  \), respectively. The procedures 
to implement the {\sc CNot} gate
is described as follows. The laser field L prepares the target bit in \( \left| g\right\rangle  \) or 
\(\left|  e\right\rangle  \);  a one bit Hadamard transform is 
applied to the target qubit  by the first Ramsey zone R\( _{1}^{\phi} \); 
then the two-bit P-gate is realized by the off-resonant atom-field interaction in C and the second 
Hadamard transform is realized by R\( _{2}^{\theta} \). Finally the atom is
detected  simultaneously specifying the atomic and field states. The effective unitary operator related to the evolution of the atom-field in cavity C
entangled state, due to the sequential interaction of the atom with the field in
$R_{1}^{\phi}$, C and $R_{2}^{\theta}$ is given by
\begin{equation}
\label{a4}
U(\phi,\theta)= U_{2}^{\theta} \exp\left[i\mu a^{\dagger}a  \sigma^+\sigma^-\right] U_{1}^{\phi}
\end{equation}
where $U_{1}^{\phi}$ and $U_{2}^{\theta}$ are the unitary operators related to the
evolution of the joint state in $R_{1}^{\phi}$ and $R_{2}^{\theta}$,
respectively. In eqn. (\ref{a4}), $\mu=\Omega_{2}t$, where $t$ is the time 
interval for the off-resonant interaction.  Proceeding through the 
immediate states generated by the atomic passing through each of the cavities it 
is easy to show, for \(\phi=\pi\) and \(\theta=0\), the following table
\begin{center}
\begin{tabular}{lcccccc}\\
Input& $R_{1}^{\phi}$& & C &  & $R_{2}^{\theta}$& Output \\ \tableline\tableline
\(\left| g\right\rangle\otimes \left|0\right\rangle_{L} \) &$\rightarrow$& \(\frac{1}{\sqrt{2}}\left( \left|
g\right\rangle -\left| e\right\rangle \right)\otimes \left|0\right\rangle_{L}\) &$\rightarrow$& \(\frac{1}{\sqrt{2}}\left( \left| g\right\rangle - \left| e\right\rangle\right)
\otimes \left|0\right\rangle_{L} \) &$\rightarrow$& \(\left| g\right\rangle
\otimes \left|0\right\rangle_{L} \) \\
\(\left| e\right\rangle\otimes \left|0\right\rangle_{L} \)&$\rightarrow$ &\(\frac{1}{\sqrt{2}}\left( \left|
e\right\rangle +\left| g\right\rangle \right)\otimes
\left|0\right\rangle_{L}\)&$\rightarrow$& \(\frac{1}{\sqrt{2}}\left( \left| e\right\rangle + \left| g\right\rangle\right)
\otimes \left|0\right\rangle_{L} \) &$\rightarrow$& \(\left| e\right\rangle
\otimes \left|0\right\rangle_{L} \) \\
\(\left| g\right\rangle\otimes \left|1\right\rangle_{L} \) &$\rightarrow$& \(\frac{1}{\sqrt{2}}\left( \left|
g\right\rangle -\left| e\right\rangle \right)\otimes \left|1\right\rangle_{L}\)
&$\rightarrow$& \(\frac{1}{\sqrt{2}}\left( \left| e\right\rangle + \left| g\right\rangle\right)
\otimes \left|1\right\rangle_{L} \) &$\rightarrow$& \(\left| e\right\rangle
\otimes \left|1\right\rangle_{L} \) \\
\(\left| e\right\rangle\otimes \left|1\right\rangle_{L} \) &$\rightarrow$&\(\frac{1}{\sqrt{2}}\left( \left|
e\right\rangle +\left| g\right\rangle \right)\otimes \left|1\right\rangle_{L}\) &$\rightarrow$& \(\frac{1}{\sqrt{2}}\left( \left| g\right\rangle 
- \left| e\right\rangle\right)
\otimes \left|1\right\rangle_{L} \) &$\rightarrow$& \(\left| g\right\rangle
\otimes \left|1\right\rangle_{L} \) \\
\tableline\\
\end{tabular}
\end{center}
which verifies the standard {\sc CNot} truth-table.

 Above we have discussed a setup where the atoms encode the target qubit and the cavity
field mode encodes the control qubit. Nevertheless, it is also possible 
to proceed with atoms responsible by both the control and target qubit. 
In this second case, the state of the control atom must be transferred
to the cavity C and with a proper selection of the cavity state (to what we
address to the next section) the procedure for implementing the {\sc CNot} 
gate follows as above. After the second atom, which encodes the target qubit
interaction in the process described above, a third atom is sent across the system to 
read the cavity state in a process similar to the scheme already proposed 
by Sleator and Weinfurter\cite{sleator}.
 To envisage a quantum network, {\it i.e.}, the interconnection of quantum gates,
the carriers of qubits between gates can be achieved by atoms transferring the
state of one cavity to another \cite{sleator}, or even by the coupling of these cavities by
superconducting wave-guides which can be responsible by an exchange of states
\cite{meu2} between two gates.

 It is important for this proposal to include  a brief discussion of 
the realistic parameters. We first note that an atom crosses the cavity in a time of order of 10\( ^{-4} \)
s, which is well below the relaxation time of the field inside C (typically 
of the order of 10\( ^{-3} \)-10\( ^{-2} \)s for Niobium superconducting 
cavities \cite{paulo}) and below the atomic spontaneous emission 
time of (3\( \times 10^{-2} \)s) \cite{haroche,brune}.
Therefore, the limits considered in that proposal must be far away from the
problematic limits found in those
experiments. 

Our entire proposal for encoding a {\sc CNot} gate discussed here is reliant on 
being able to generate the zero ($\left| 
0\right\rangle_{L}$) and one ($\left| 1\right\rangle_{L}$) logical 
states. For this reason we address in Section (\ref{Initial 
Conditions}) a strategy for guaranteeing the exact choice of the initial 
cavity field state. Without such a strategy, the logical states can 
only theoretically be generated with a $50\%$ probability. More 
explicitly there is a $50\%$ probability that the $\left| 
0\right\rangle_{L}$ state actually contains only even photon number 
states  and a $50\%$ probability that it contains only odd photon number 
states.

%
\section{Initial Conditions for the Control Bit}\label{Initial Conditions}
%
Our generation of the {\sc CNot} gate outlined in the previous section relies on our ability to be able to generate the coherent logical state 
encoding with a high degree of certainty.
The initial state of the control bit  (the field state of the cavity) has 
to be prepared with a probability greater than 50\% 
as usually occurs in the preparation of superposition field
states by Rydberg atoms. The state of the field in the cavity is
$\left|0\right\rangle_{L}$ or $\left|1\right\rangle_{L}$ conditioned by the
measurement of the atomic $\left|g\right\rangle$ or $\left|e\right\rangle$ state 
in the process of generation of superposition states.  Such a scheme 
is analogous to the depicted in fig.(1), however here we have $\theta=\pi$ in the second Ramsey zone and for 
the initial cavity state a coherent one, considering that the atom was prepared 
in the $\left|e\right\rangle$ state.
Let us suppose we are interested in preparing the state $\left|0\right\rangle_{L}$
for the control bit. If the atomic state \(\left|e\right\rangle\) was detected,
then our  scheme would  have failed. For it to succeed we have to apply a process conditioned to the
measurement of the atomic $\left|e\right\rangle$ state to guarantee the flip of
the cavity field state from $\left|1\right\rangle_{L}$ to
$\left|0\right\rangle_{L}$. Analogously we have to apply a process conditioned to the
measurement of the atomic $\left|g\right\rangle$ state to guarantee the flip of
the cavity field state from $\left|0\right\rangle_{L}$ to
$\left|1\right\rangle_{L}$ if we are interested in prepare the control bit in
the $\left|1\right\rangle_{L}$ state. 

First noting the fact that an atom interacting resonantly with the field in C, with a controlled velocity, can
exchange a single photon and regarding that a single photon emission by the cavity field causes
\begin{eqnarray}
\label{b22}
a \left|0\right.\rangle_{L} &=&\alpha\frac{N_{-}}{N_{+}}\left|1\right\rangle_{L} \approx \alpha \left|1\right\rangle_{L}
\;\;\;\;\;({\rm \alpha\;\;large}) \\
a\left|1\right.\rangle_{L} &\approx& \alpha \left|0\right\rangle_{L}.
\end{eqnarray}

We can now formulate an atomic feedback scheme that operates 
whenever the atomic detector clicks, if
we are interested in the control  qubit $\left|0\right\rangle_{L}$ or
$\left|1\right\rangle_{L}$. In fact this process is very similar to the stroboscopic
feedback proposed by Vitali {\it et al.}\cite{milburn} for the suppression of
decoherence of superposition field states. Of course we do not need a stroboscopic action, but just one event
conditioned to the atomic state measurement.

The scheme proposed is depicted schematically in fig.(2), where B$_{2}$ is a source of atoms
which are tuned in resonance with the field in C by the Stark shift conditioned
to the atomic state measurement made in the ionization zones D$_{e}$ or D$_{g}$.
The resonant atom-field interaction is given by the Hamiltonian

\begin{equation}
\label{a421}
H_{I}=\hbar \Gamma \left( a\sigma ^{+}_{f}+a^{\dagger }\sigma ^{-}_{f}\right) 
\end{equation}
where $\Gamma$ is the coupling constant between the field and 
atomic variables. Here $\sigma ^{+}_{f}$ and $\sigma ^{-}_{f}$ are rising and
lowering operator for the feedback atom.  
If the feedback atom is prepared in the state $\left|e\right\rangle$ then the field
state is given by
\begin{eqnarray}
\rho^{g\choose e} _{f} 
&=& \cos (\Gamma \tau\sqrt{a^{\dagger}a+1})\;\rho^{g\choose e}_{C} \;\cos (\Gamma
\tau\sqrt{a^{\dagger}a+1})+ a^{\dagger} \;\frac{\sin (\Gamma
\tau\sqrt{a^{\dagger}a+1})}{\sqrt{a^{\dagger}a+1}}\;\rho_{C}^{g\choose e} \;\frac{\sin (\Gamma
\tau\sqrt{a^{\dagger}a+1})}{\sqrt{a^{\dagger}a+1}}\; a.\label{a422} 
\end{eqnarray}
where $\rho^{g\choose e}_{C}$ is the density operator associated with the field state in C 
before the feedback action.  Here $\rho^{g} _{f}$ ($\rho^{e} 
_{f}$) explicitly is the ground (excited) state density operator.  
$\tau$ is the time of interaction of the feedback atom with the field. 

As a measure of the field state in the cavity a second atom is sent through the
 setup and again measured in D$_{g}$ or D$_{e}$ \cite{davidovich}. The conditional 
probability $P^{g \choose e}(T)$ that the second atom  will be
detected in the \( \left| g\right\rangle  \) or \( \left| e\right\rangle  \)
state, at the time $T$ after detection of the first atom follows
\begin{eqnarray}
P^{g \choose e}(T) 
& = & \frac{1}{2}\left\{ 1\pm \frac{1}{1+\cos \varphi \mbox{e}^{-2\left| \alpha \right| ^{2}}}
 \left[ \mbox{e}^{-2\left| \alpha \right| ^{2}\mbox{e}^{-\gamma T}}
 +\cos \varphi \mbox{e}^{-2\left|
  \alpha \right| ^{2}\left( 1-\mbox{e}^{-\gamma T}\right) }
   \right] \right\} ,\label{g6} 
\end{eqnarray}
conditioned to \( \varphi =0 \) [\( \pi  \)] if the first atom is detected in the \( \left| g\right\rangle _{1} \)
[\( \left| e\right\rangle _{1} \)] state and to the signal + [-] for the second
atom be detected in the \( \left| g\right\rangle _{2} \) [\( \left| e\right\rangle _{2} \)]
state. For the computation of Eq. (\ref{g6}) at time T we have included the relaxation
of the field state due to dissipation. Considering a reservoir at zero
temperature, this state is now given by
\begin{eqnarray}
\rho _{C}^{g\choose e}(T) & = & \frac{1}{N^{2}_{\pm}}\left\{ |\alpha \mbox{e}^{-\gamma
T/2}><\alpha \mbox{e}^{-\gamma T/2}|+|-\alpha \mbox{e}^{-\gamma T/2}><-\alpha
\mbox{e}^{-\gamma T/2}|\right. \nonumber \\
&\pm&\left. \mbox{e}^{-2|\alpha |^{2}(1-\mbox {e}^{-\gamma T})} \left[ |-\alpha \mbox{e}^{-\gamma T/2}><\alpha
 \mbox{e}^{-\gamma T/2}|+|\alpha \mbox{e}^{-\gamma T/2}><-\alpha
 \mbox{e}^{-\gamma T/2}|\right] \right\}.\label{a131} 
\end{eqnarray}
where $\gamma$ is the relaxation constant of the field. By analyzing Eq.(\ref{g6}) 
 we observe that if the second atom is detected instantaneously
after the first one ($\gamma T\ll 1$) then
\begin{equation}
\label{g7}
P^{g\choose e}(T)=
\frac{1}{2}\left[ 1\pm \cos \varphi \right],
\end{equation} 
again with $\varphi=0$ $ (\pi)$. This gives the conditional probability of detection of the first atom in 
\( \left| e\right\rangle _{1} \) and the second atom in \( \left| 
e\right\rangle _{2}\) as $P(e,e)\equiv P^{g \choose e}|^{\pi, -}=1$ and the probability of detection of the 
first in \( \left|g\right\rangle _{1} \)and the second in \( \left| e\right\rangle _{2} \) as
$P(g,e)\equiv P^{g \choose e}|^{0, -}=0$. Analogously $P(g,g)\equiv
P^{g \choose e}|^{0,+}=1$ and $P(e,g)\equiv P^{g \choose e}|^{\pi, +}=0$. 
This is a signature of the measurement for the state in which the cavity field was
prepared. It presents our undesired results $P(g,e)$ and $P(e,g)$ 
equal to zero, that is no probability of them occurring. However if the feedback loop is taken in to account in the calculation of the
probabilities $P(g,e)$ and $P(e,g)$ then, instead of using $\rho^{g\choose e}_{C}$ 
in Eq. (\ref{g6}) we must use $\rho^{g\choose e}_{f}(T+\tau)$ from 
Eq. (\ref{a422}), where $T'=T+\tau$. Substituting Eq. (\ref{a131}) for the 
field relaxation into Eq. (\ref{a422}) it follows
\begin{eqnarray}
P^{g\choose e}_{f}(T+\tau) & =& \frac{1}{2}\left\{1\pm\frac{2}{N^{2}} \mbox{e}^{-\left|\alpha\right|^{2}}
 \sum_{m} \frac {\left(-\left|\alpha\right|^{2}\mbox{e}^{-\gamma T}\right)^{m}}
 {m!} \left[ 1+ (-1)^{m} \cos \varphi
 \; \mbox{e}^{-2\left|\alpha\right|^{2}\left(1-\mbox{e}^{-\gamma T}\right)}\right]
 \cos\left(2\Gamma \tau \sqrt{m+1}\right)\right\},\label{gg9}
\end{eqnarray}
which accounts for the conditional probability of detection of the second atom
in the state \( \left| g\right\rangle _{2} \) [\( \left| e\right\rangle _{2} \)]
if the first atom was detected in the state \( \left| e\right\rangle _{1} \) 
[\( \left| g\right\rangle _{1} \)].
In Fig. (3) we show the respective four conditional probability of atomic detection,
 $P(e,e)$ and $P(g,g)$ without feedback and  $P_{f}(e,g)$ and $P_{f}(g,e)$ considering the
 feedback loop for some values of $\Gamma \tau$. This shows the feasibility of
 the feedback to control the initial state of the cavity. The figure is plotted
 until $\gamma T=1$ since there is no reason to consider times longer than this
 once the decoherence of the state has already taken place. In fact the scale of
 time to be taken into account in figures 3(c) and 3(d) is $T'=T+\tau$, the time
 interval after the detection of the first atom plus the time interval of the
 feedback atom. In these figures the continuous solid line represents the absence of feedback. 
 As can be seen there is an optimum value
 for the feedback process at $\Gamma \tau = \pi/6$ which gives a 93\% of chance for the
 cavity field qubit to be prepared in the right state. It also must be noted that an optimal value
 is possible only when the feedback atom is sent instantaneously after the
 click of the respective detector. The performance of the setup decreases
 considerably when a time delay exists, as can be observed in Figs. 3(c) and
 3(d) for $\gamma T'\ge 0.1$. The limit of those curves around 0.5  means that the field
 state already decohered, and so there is 50\% of chance again for generation of
 the \( \left| 0\right\rangle _{L} \) or \( \left| 1\right\rangle _{L} \)
 states. For $\gamma T'> 1.0$ (not shown in the figures) the effect of
 dissipation  implies amplitude damping. The field asymptotically tends to be in a
 vacuum state, and when this occurs is easily shown through Eq. (\ref{gg9}) that the
 second atom will always be detected in  the 
 \(\left|g\right\rangle\) state. With the
 feedback it tends to always be detected in \(\left|e\right\rangle\) 
 state.
%

\section{Efficiency and Sources of Error}
 This section discusses in detail the advantages and disadvantages of
encoding qubits in superposition states instead of number states of only one
and zero photon and the effect of dissipation on these.  
As is already well known for cavity QED experiments the dominant source of error  
that will affect the implementation of quantum logic elements is cavity 
damping. Since the cavities are not isolated, when the states \(\left| 0\right\rangle_{L}\) or 
\(\left|1\right\rangle_{L}\) are constructed, the presence of dissipative
effects will alter the free evolution of the cavity field state introducing amplitude damping
as well as coherence loss. The  zero temperature master equation describing 
the  bosonic damping is simply
\begin{equation}
\label{f0}
\frac{d\rho}{dt}=\frac{\gamma}{2}\left(2a\rho a^{\dagger}-a^\dagger a \rho-
\rho a^\dagger a\right),
\end{equation}
and its solution for any initial state can be written as \cite{milburn}
\begin{equation}
\label{f01}
\rho(t)=\sum_{k=0}^{\infty}\Upsilon_k(t)\rho(0)\Upsilon_k^\dagger(t),
\end{equation}
where
\begin{equation}
\label{f02}
\Upsilon_k(t)=\sum_{n=k}^\infty\sqrt{n\choose{k}}\left(e^{-\gamma t}\right)^{(n-k)/2}
\left(1-e^{-\gamma t}\right)^{k/2} \left| n-k\right\rangle\left\langle
n\right|.
\end{equation}
We are interested in the effect of dissipation on the information encoded in the
qubits. For that we will consider first a superposition of Schr\"odinger cats
qubits and thereafter a superposition of one and zero photon number states qubit
encoding.

 The action of a single decay event \(\Upsilon_{1}\) on the state 
\begin{equation}
\label{fa4}
\left|\psi_1\right\rangle =
E_1\left|0 \right\rangle_{L} +E_2\left| 1\right\rangle_{L},
\end{equation}
leads to
\begin{equation}
\label{f05}
\Upsilon_{1}\left|\psi_1\right\rangle =\alpha \left(1-e^{-\gamma
t}\right)^{1/2} e^{-|\alpha|^2\left(1-e^{-\gamma
t}\right)/2}\left(E_1\frac{N_-}{N_+}\left|\tilde{1} \right\rangle_{L}
+E_2\frac{N_+}{N_-}\left| \tilde{0}
\right\rangle_{L}\right),
\end{equation}
that is, a simple bit-flip occurs. Here $\left| \tilde{0}
\right\rangle_{L}\equiv \frac{1}{N_{+}}\left( \left|
e^{-\gamma t/2}\alpha \right\rangle +\left| -e^{-\gamma t/2}\alpha \right\rangle \right)$ and $\left| \tilde{1}
\right\rangle_{L}\equiv\frac{1}{N_{-}}\left( \left|
e^{-\gamma t/2}\alpha \right\rangle -\left| -e^{-\gamma t/2}\alpha \right\rangle \right)$, account for the amplitude damping. A simple 
unitary process will transform Eq. (\ref{f05}) back to Eq. (\ref{fa4}), meaning 
the reversibility of the process. Under a double decay event \(\Upsilon_{2}\),
\begin{equation}
\label{f06}
\Upsilon_{2}\left|\psi_1\right\rangle =\alpha^{2} \left(1-e^{-\gamma
t}\right)e^{-|\alpha|^2\left(1-e^{-\gamma t}\right)/2}\left(E_1\left|\tilde{0}
\right\rangle_{L} +E_2\left|\tilde{1}
\right\rangle_{L}\right),
\end{equation}
which is exactly our initial state but with amplitude damping. This special
superposition is invariant under even number of decay
events. This fact brings one important information about these 
states. However a single decay event, \(\Upsilon_{1}\) on the 
Fock superposition state
\begin{equation}
\label{f1}
\left|\psi_2\right\rangle =F_1\left|0 \right\rangle +F_2\left| 1 \right\rangle,
\end{equation}
leads to
\begin{equation}
\label{f2}
\Upsilon_{1}\left|\psi_2\right\rangle =F_2\left(1-e^{-\gamma
t}\right)^{1/2}\left| 0 \right\rangle.
\end{equation}
No unitary operation can recover (\ref{f1}) indicating the irreversibility of the process.
 This means that in one photon state information processing
schemes, one single photon decay is fatal, since there is no way in which the
 resulting error  can be corrected once it occurs. However for qubits
consisting of superpositions of odd and even number states, one decay event
cause a bit-flip, which could be, in principle be corrected.
So here, we classified two different kind of error
arising from dissipation, one impossible to be corrected (called irreversible
error) and the other a bit-flip which can be corrected (reversible error) by unitary 
processes\cite{munro,knill}.

There is a number of error  correction schemes  that protect the quantum 
information against single errors. 
As we have mentioned previously a 
spontaneous emission error  for the Schr\"odinger cat encoding 
results in a bit-flip. It is well 
known that such errors can easily be prevented by a 
3-qubit error correction circuit \cite{Braunstein} 
(schematically depicted in Fig (4a)). This circuit is reasonably 
simple and the superposition state it produces is relatively simple. 
In fact  for an arbitrary qubit $\left|\psi\right\rangle= E_1 
\left|0\right\rangle_{L}+ E_2 \left|1\right\rangle_{L}$ the correction circuit generates 
the encoded superposition state
\begin{eqnarray}
\left|\psi\right\rangle&=&E_1 \left|000\right\rangle+E_2
\left|111\right\rangle.	
\end{eqnarray}

To protect against arbitrary error requires normally a 5-qubit error correction 
circuit \cite{Laflammeerror} (schematically depicted in Fig (4b)). 
For an arbitrary qubit $\left|\psi\right\rangle = F_1 \left|0\right\rangle + F_2
\left|1\right\rangle $, 
the correction circuit generates the superposition state
\begin{eqnarray}
\left|\psi\right\rangle &=&F_1 \left[\left|00000\right\rangle
+\left|00110\right\rangle +\left|01001\right\rangle -\left|01111\right\rangle +
\left|10011\right\rangle+\left|10101\right\rangle +\left|11010\right\rangle +\left|11100\right\rangle\right] \nonumber \\
&+&F_2 \left[\left|00011\right\rangle -\left|00101\right\rangle
-\left|01010\right\rangle -\left|01100\right\rangle -\left|10000\right\rangle
+\left|10110\right\rangle +\left|11001\right\rangle
+\left|11111\right\rangle\right].	
\end{eqnarray}
This is quite a complicated superposition state to create (as can be 
seen from the quantum circuit). The 3-qubit correction circuit 
is much simpler and hence we see the advantage of the Schr\"odinger 
cat encoding. Also, while
here we are only discussing a single gate, a reasonable quantum computer has
to be constituted of many gates. Then, if the above  5-qubit protection circuit has
to be implemented, this will become much more expensive in terms of qubits in
comparison to the 3-qubit circuit for bit-flip protection. The bit-flip
protection scheme saves 2 qubits at each needed qubit in comparison to the 5-qubit
protection circuit described above.  It does however only protect against a
specific type of error.
An unavoidable error incoming from dissipation  over superposition
states is decoherence. Let us consider the general effect of dissipation on the quantum 
coherent superposition state. At zero temperature the state of the cavity
field is described by the density operator

\begin{eqnarray}
\rho _{C}^{\pm}(t) & = & \frac{1}{N^{2}_{\pm}}\left\{ |\alpha \mbox{e}^{-\gamma
t/2}><\alpha \mbox{e}^{-\gamma t/2}|+|-\alpha \mbox{e}^{-\gamma t/2}><-\alpha
\mbox{e}^{-\gamma t/2}|\pm \mbox{e}^{-2|\alpha |^{2}(1-\mbox {e}^{-\gamma t})}\right. \nonumber \\
 &  & \left. \times \left[ |-\alpha \mbox{e}^{-\gamma t/2}><\alpha
 \mbox{e}^{-\gamma t/2}|+|\alpha \mbox{e}^{-\gamma t/2}><-\alpha
 \mbox{e}^{-\gamma t/2}|\right] \right\}.\label{a13} 
\end{eqnarray}
 We see that two characteristic times are involved in this evolution. The first one, the
\emph{decoherence time} is the time in which the pure state given by Eq.(7)
is turned into a statistic mixture
\begin{equation}
\label{a8}
\rho _{C}(t)\approx \frac{1}{2}\left\{ |\alpha ><\alpha |+|-\alpha ><-\alpha
|\right\} ,
\end{equation}
the second is the \emph{damping time} or \emph{relaxation time} of the field
\( t_{c} \) =\( \gamma ^{-1} \), the time that the dissipative effect reduces
the energy of the field leading it in to a vacuum state.

The decoherence of the field state is characterized by the \( \exp 
\left[ -2\left| \alpha \right|^{2}\left( 1-\mbox{e}^{-\gamma t}\right) \right]  \)
factor, that for short times, \( \gamma t\ll 1 \), turns to be \( 
\exp \left[ -2\left| \alpha \right|^{2} \gamma t\right]  \)
and the coherence decays with the time \( t_{d}=\left( 
2\gamma \left| \alpha \right|^{2}\right) ^{-1} \).  
Unfortunately the coherence time depends on inversely on 
$\left| \alpha \right|^{2}$ and hence the larger the $\left| \alpha 
\right|^{2}$ the smaller the coherence time. Decoherence constitutes the 
main obstacle to quantum computation \cite{dec}, since the encoding is completely based in the
purity of the field state.

The relaxation time of microwave fields in superconducting cavities is  of the order of
\(10^{-2}\) s \cite{haroche}, what means $t_{d}\approx 10^{-2}\left| \alpha \right| ^{-2}$ s. So
all the interactions involved in this proposal must consider this time and more
specifically the number of photons as critical quantities. Moreover, the initial
information encoded in the superpositions given by Eqs. (\ref{fa4}) and
(\ref{f1}) also suffer the effect of decoherence, which for $|\alpha|\approx 1$
occur at the same time for both encoding schemes. Again, decoherence prevention
schemes play a crucial role for any quantum information encoding.

One favorable point
for the superposition state encoding is that proposals for sustaining the coherence of these field states have already been considered which could be well adapted for 
 our case.  For example the stroboscopic feedback proposal of Vitali {\it et
 al.} \cite{milburn}.  This proposal is particularly appropriate here since it guarantee that
at each single decay event a feedback atom is sent through the setup restituting the coherence
and the state parity. In fact in \cite{milburn} the authors claim that the
coherence is restored but for a slightly different state. For the
proposal presented here what is important is not the original superposition of
states, but the original parity of the state, if it was originally a
superposition of even or odd photon number states. 
It is important to emphasize the experimentally critical values for the physical
elements involved. The time of flight of the atom across the setup (10\( ^{-4} \)
s), the relaxation time of the field (10\( ^{-3} \)-10\( ^{-2} \)s for Niobium superconducting cavities)
and the atomic spontaneous emission time (3\( \times 10^{-2} \)s)
\cite{haroche,paulo}.

%
%
\section{Conclusion}

In conclusion we have presented a feasible scheme to encode the
{\sc CNot} quantum gate, based on a field superposition of states. These states have been already generated in superconducting microwave cavities 
which constitute a system almost dominated by the current technology \cite{haroche}. The proposal 
 here to encode the {\sc CNot} gate based on a superposition of
states is less susceptible to irreversible errors due to dissipative effect imposed by the environment
than number states \cite{sleator,domokos}. The generation of these kind of states is dependent of a conditional
measurement giving a random assignment of initial control bits, which would be
useless if no further process is considered. Hence we propose a conditional
feedback scheme, which guarantees that the initial control bit is prepared in
the required
state. Once the amplitude damping of a coherent state (at zero temperature)
still constitutes a coherent state the method proposed works until the inevitable
effect of decoherence takes place. For that a reset of the qubits must be done
after a time of the order of the time of the decoherence or a coherence control
scheme \cite{milburn} must be applied. The reset process is done
repeating the process here described.

The state of a logic unit can be transferred to another logic unit 
(if the time of decoherence is respected), constituting a sort of quantum memory
circuit \cite{memdavid}. That can be attained by the proper choice of atomic interactions between
atoms and the field in the microwave cavity or even by the direct photonic
process of coupling two cavities by a superconducting wave-guide, which permits
an exchange of information (exchange of states) between the coupled units
\cite{meu2,coupcavity}. This problem is of fundamental importance for the
engineering of quantum networks\cite{zoller2}. The
major sources of error here are the loss of coherence of the field state and the
control bit-flip due to the dissipative effect. Analysis of these kind of errors
on quantum networks constituted by the basic element here described is left
for further investigation. 

\acknowledgments{MCO thanks the Funda\c{c}\~ao de Amparo \`a Pesquisa do Estado
de S\~ao Paulo (Brazil) for financial support while WJM acknowledges the support of the Australian Research Council.
}
%

%
%
\begin{figure}[tbp]
\caption{Experimental setup for a {\sc CNot} gate. Here B is a source of atoms, L is a
laser field which prepares the atomic state, C is a superconducting microwave
cavity, R$^{\phi}_{1}$ and R$^{\theta}_{2}$ are Ramsey zones and D is an
ionization zone atomic detector, while Ss are classical microwave sources. The state of the field in C encodes the control bit and the atomic
state, the target bit.}
\label{fig1}
\end{figure}
\begin{figure}[tbp]
\caption{Experimental feedback setup for the control of the initial state of the field in
C. Here B$_{1}$ and B$_{2}$ are atomic sources of atoms, L is a
laser field that prepares the atomic state, C is a superconducting microwave
cavity, R$^{\phi}_{1}$ and R$^{\theta}_{2}$ are Ramsey zones, D$_{e}$, D$_{g}$ are
ionization zone atomic detectors and Ss are classical microwave sources. Once the undesired state is measured by D$_{e}$ or D$_{g}$ the
B$_{2}$ source of atoms is turned on and a resonant interacting atom flips the state of the
field in C.} 
\label{fig2}
\end{figure}

\begin{figure}[tbp]
\caption{Conditional probability of detection for the second 
atom being in a particular state given the result of a measurement on 
the first atom. Figure a) shows the probability of obtaining the 
second atom in the  excited state $\left|e\right\rangle_{2}$ given the first atom was 
measured in the $\left|e\right\rangle_{1}$ state. Figure b) shows the probability of obtaining the 
second atom in the   ground state $\left|g\right\rangle_{2}$ given the first atom was 
measured in the $\left|g\right\rangle_{1}$ state. Figure c)  gives the 
conditional probability of detecting the second atom in the  
$\left|g\right\rangle_{2}$ state given the first atom was 
measured in the $\left|e\right\rangle_{1}$ when the feedback action 
is taken into account. Here \(T'=T+\tau\). Figure d)  gives the 
conditional probability of detecting the second atom in the  
$\left|e\right\rangle_{2}$ state given the first atom was 
measured in the $\left|g\right\rangle_{1}$ when the feedback action 
is taken into account. Again \(T'=T+\tau\). In Figures c) \& d) the 
solid line represents the absence of feedback.
}
\label{fig3}
\end{figure}

\begin{figure}[tbp]
\caption{Three (a) and Five (b) qubit error correction circuits. The 
3-qubit circuit correct bit-flip errors while the  5-qubit circuit 
correct arbitrary errors.}
\label{fig4}
\end{figure}

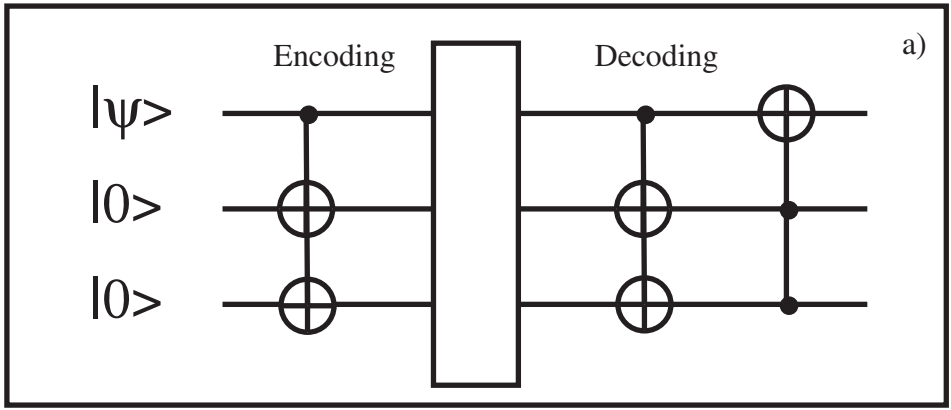
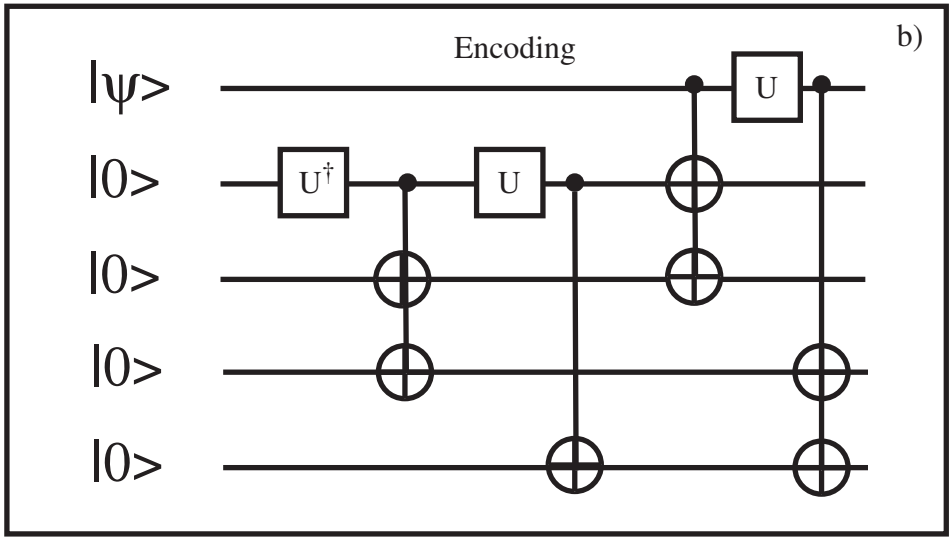
\end{document}